\documentclass[acmsmall]{acmart}

\AtBeginDocument{%
  \providecommand\BibTeX{{%
    \normalfont B\kern-0.5em{\scshape i\kern-0.25em b}\kern-0.8em\TeX}}}

\setcopyright{acmcopyright}
\copyrightyear{2023}
\acmYear{2023}
\acmDOI{10.48550/arXiv.2307.11060}

\acmConference[SE4RS'23]{Software Engineering for Research Software}{July 24,
  2023}{Portland, OR}
\acmPrice{15.00}
\acmISBN{978-1-4503-XXXX-X/18/06}

\begin{document}

\title{The Changing Role of RSEs over the Lifetime of Parsl}

\author{Daniel S. Katz}
\authornote{All authors contributed equally to this research.}
\email{d.katz@ieee.org}
\orcid{0000-0001-5934-7525}
\affiliation{%
  \institution{University of Illinois Urbana Champaign}
  \streetaddress{1205 W Clark St}
  \city{Urbana}
  \state{Illinois}
  \country{USA}
  \postcode{61801}
}

\author{Ben Clifford}
\authornotemark[1]
\email{benc@hawaga.org.uk}
\orcid{0000-0001-6397-7239}
\affiliation{
  \institution{CQX Limited}
  \country{United Kingdom}
}

\author{Yadu Babuji}
\authornotemark[1]
\email{yadunand@uchicago.edu}
\orcid{0000-0002-9162-6003}
\author{Kevin {Hunter Kesling}}
\authornotemark[1]
\email{hunteke@uchicago.edu}
\orcid{0000-0003-2828-4511}
\author{Anna Woodard}
\authornotemark[1]
\email{annawoodard@uchicago.edu}
\author{Kyle Chard}
\authornotemark[1]
\email{chard@uchicago.edu}
\orcid{0000-0002-7370-4805}
\affiliation{%
  \institution{\\University of Chicago}
  \streetaddress{5730 South Ellis Avenue}
  \city{Chicago}
  \state{Illinois}
  \country{USA}
  \postcode{60637}
}

\renewcommand{\shortauthors}{Katz, Clifford, Babuji, Hunter Kesling, Woodard, and Chard}

\begin{abstract}
  This position paper describes the Parsl open source research software project and its various phases over seven years. It defines four types of research software engineers (RSEs) who have been important to the project in those phases; we believe this is also applicable to other research software projects.
\end{abstract}



\received{26 May 2023}

\maketitle

\section{Introduction}

Parsl~\cite{babuji19parsl, parsl-github} is an open source software package and community that enables Python programmers to parallelize and scale their work to the largest computers. Programmers can wrap Python functions and executables into higher-level programs, where those functions and executables can be run concurrently and executed remotely on different types of systems, such as laptops, clusters, clouds, and high-performance computing (HPC) clusters.
Parsl serves as a basis for novel computer science research related to parallel and distributed computing as well as a practical tool for users to write parallel and distributed programs.

Parsl started as an idea in 2016 and was initially supported by an NSF SI2 award until 2022, which supported 2-4 people/FTEs per year. Parsl is now supported by NSF and Chan Zuckerberg Initiative (CZI) awards focused on sustainability, 
support from external projects that make use of Parsl, 
volunteer contributions from users, and from 
a company who makes use of Parsl.

As the project has evolved from a single research software engineer (RSE) who developed the initial prototype 
to a community (with 73 contributors currently), the
project has undergone significant changes. These changes
include in the types of contributors, development and maintenance processes used, community culture, and 
funding sources and models used to support the project.

\section{Project evolution and challenges}
As Parsl has transitioned from a single-developer prototype
towards a sustainable project with a community of developers, users, and facilitators, we 
have faced a range of challenges, including funding 
and community roles.

During the initial Parsl project work, the resources that were provided and volunteered were useful for adding features and supporting a limited number of use cases, but were not sufficiently coordinated or aligned to fully support Parsl’s core needs over multiple years. Some of our current work is aimed toward rectifying this and making Parsl more community-maintainable in the future.

\subsection{Funding}
Today, a combination of sources support Parsl development, including a 2-year NSF award aimed at increasing sustainability, a CZI award aimed at impact in the life sciences, in-house funding at U Chicago and U Illinois, funding from projects and collaborations that depend on Parsl and help support it, and volunteer (in-kind) effort from groups that develop tools that use Parsl or add to its capabilities, students who come across Parsl and find it useful, research projects that extend Parsl in various ways, and companies that use Parsl in their services. 

A key example of funding from an external project is that a portion of Parsl's maintenance and development work has been supported by the Dark Energy Science Collaboration (DESC)~\cite{Abolfathi_2021} during the latter part of the SI2 award and continuing through today. They've funded 25 percent of the main Parsl maintainer over the last several years, and the bulk of the work they've funded has been improving the Parsl code, driven by close application integration with a real application. 

Satisfying the needs of these very different stakeholders
has been challenging. Research funding enables exploration
of new ideas, often without clear use cases or identified users who would make use of these contributions. Project funding (e.g., from DESC) is focused on addressing specific 
community needs, enhancing Parsl for specific use cases and resolving specific bugs. Sustainability-focused work, industry-supported development, and contributions from computing centers that support Parsl have equally specialized needs. We see conflicting goals from these
different projects (and their communities) as well as 
differences in the  kinds of developers who are involved, and what they want to accomplish, both for the project and for themselves.

\subsection{Community governance and roles}

To become sustainable (able to assemble the resources to continue to support the project), we are transitioning to a community-governed and supported open source project, with income handled and distributed by a 501(c)(3) (e.g., NumFOCUS) under the direction of an elected Parsl Coordination Committee. This will allow Parsl to become independent of its current leaders and their institutions, and provides a way to distribute funding to the best-suited person for each activity.

Our current needs (e.g., maintenance, outreach, and support) differ from earlier in the project, leading to a need for new types of contributions. We need to maintain the Parsl codebase, including responding to issues, reviewing contributed code, and adding additional tests to improve code quality. This leads us to develop minimal requirements on contributed code, starting with a pre-coding discussion and including plans for future maintenance and support of contributed code.

A significant fraction of developer time on Parsl
is now spent responding to user queries (e.g., via an open Slack channel) and supporting deployment on different computing resources. As the community grows and diversifies, the range of use cases and the range of challenges also grow. Members of the community can support outreach activities (e.g., presenting tutorials, hosting summer students, developing training materials for various domains),
coordinating the yearly user meeting, applying for funding, advertising success stories in blogs, and managing social media. 

How these needs are generally addressed depends on the types of community members involved, which changes depending on a project's goals and most importantly, its means of support. 

\section{Developer Types}

During our work in Parsl, we have identified four types of programmer, often a RSE, each of which has characteristics that make them well-suited for various research software project phases:

\begin{itemize}

\item {Research prototyper}: prototypes ideas very quickly. This type of developer is essential to bootstrap a new research software project, to develop and test the initial ideas, focusing on what the software can do more than how users will use it. They also might take shortcuts at this point that could harm the later sustainability of the project, adding technical debt.

\item {User/developer}: a scientist or disciplinary researcher who also adds features relevant to their work. They typically are not deeply steeped in the development side of this project, but they write code elsewhere for their real job (research), and they are a power user of the software who can understand bugs and fix them. 
They are primarily focused on their own usage of the software, and again, they may take shortcuts that harm the project's future sustainability. 
They are important to have involved in all but the initial stage of the project.

\item {Software developer}: focuses on the research software itself and its users. They are dedicated to making the software as useful as possible to its users, and often also seek to make it clean and beautiful, increasing simplicity, compatibility, and maintainability in the future, and reducing technical debt that arises in earlier development. They are important to have involved in all but the initial stage of the project (where process may impede fast innovation).

\item{Collaborating developer}: responsible for a collaborative project, and wants their software to work with or be ``part of'' Parsl. In their own project, they may be any of the above types. They may also follow a different coding and software engineering style, potentially leading to integration and culture challenges. Defining the interface to these developers is a key challenge for sustainability, as ideally, they will become committed to Parsl's success as important to their own project's. While not required, this role is most helpful after the project has become initially proven and has demonstrated some success.

\end{itemize}

\section{Lessons and Future Work}
The existence of different types of RSEs and their utility during different phases of Parsl, particularly when moving from a project funded by a grant to one supported by a mix of sources, emerged during the project. Future work could study different developers to understand their preferred type (style of work), and other 
research software projects to see if our experiences are shared. 

\begin{acks}
This work was supported by NSF (1550588, 2209919, 2209920) and the Chan Zuckerberg Initiative.
\end{acks}

\bibliographystyle{ACM-Reference-Format}
\bibliography{software}

\end{document}